# Exponential Convergence of Piecewise-Constant Parameters Identification under Finite Excitation Condition

Anton Glushchenko, *Member, IEEE*, Vladislav Petrov, and Konstantin Lastochkin

*Abstract*—A problem of identification of piecewise-constant unknown parameters of a linear regression equation (LRE) is considered. Such parameters change their values over the interval of the regressor finite (rather than persistent) excitation. To solve it, the previously proposed I-DREM procedure and the integral-based filtering method with the exponential forgetting and resetting are improved: the integral of the filter equations is taken over the finite time intervals, which belong to the finite excitation time range. This allows one to obtain an exponentially bounded identification error over the excitation interval, and, when the LRE parameters are constant outside such interval, to provide exponential convergence of the identification error to zero. In addition, the applied method of the regressor excitation normalization makes it possible to obtain the same rate of convergence of the parameter error for the regressors of various amplitudes. The stability and all the above-mentioned properties are proved for the derived identification method. The results of numerical experiments (including the case of the noise-contaminated regressor and output measurements) fully support the obtained theoretical results.

## I. INTRODUCTION

In recent years, the problem of quality improvement of the linear regression equation (LRE) parameters identification has attracted considerable attention of the control science community. The main studies in this research area are aimed at relaxation of the regressor persistent excitation (PE) condition to provide exponential convergence of the estimation (identification) error of the unknown LRE parameters to zero [1].

Most of the methods, which have been proposed in the literature [2-12] to relax this condition, can be divided into 3 approaches [2-4] based on identification schemes with the finite-time convergence, 2) the ones [5-12], which propose different algorithms of the regressor pre-processing, and 3) the method [13] to generate a new initially exciting regressor.

In this paper, let the drawbacks of the existing algorithms of the regressor pre-processing be considered in more detail. Among them, it is necessary to highlight the Dynamic Extension and Mixing (DREM) procedure [12], and various versions [5-11] of the integral-based filtering of the LRE.

In comparison with other pre-processing methods, the originality of the DREM procedure [12] lies in the generation of a scalar regressor, which allows one, firstly, to provide monotonicity of the unknown parameter estimation process, and, secondly, to obtain a new condition of the regressor non square-integrability for the asymptotic convergence of the unknown parameters estimation error to zero. However, this condition, although being strictly weaker than the PE one, is still strong enough for many practical applications. Therefore, in the literature, the following methods are proposed to relax the convergence conditions of DREM: the FCT-D estimation law [2], the approach to generate a new scalar exciting regressor [13], and the I-DREM procedure [11].

The general concept of the LRE integral-based filtering methods [5-11] is the idea of minimization not only of the current identification error but also of its integral by an estimation method. This allows one to continue the gradient search for the unknown parameters even when the regressor initial excitation is vanishing. To obtain the integral of the estimation (identification) error, the initial LRE is filtered using various integral-based filters to obtain a non-vanishing filtered regressor from the initial one. All such methods [5-11] can be conventionally classified by the type of the new condition, which is necessary to ensure the convergence of the estimation process: either initial (IE) or finite (FE) excitation one.

The above-mentioned filtering methods, which relax PE to IE, use the integral-based filters, which are sensitive to the data on the LRE only over some initial time interval. Examples of such filters are an open-loop integrator with a switching algorithm [14], an integrator with the exponential input decay [11], etc. In turn, the methods [8-10], which relax PE to FE, are sensitive to the data on the LRE over an arbitrary time interval. Usually, such approaches use non-stationary aperiodic filters, which time constant changes its value according to a special analytical law.

The main problem of the most known integral-based filtering methods [5-11] is the inaccurate identification of the time-varying LRE parameters. This problem is caused by the fact that the integral of the estimation error, in this case, is calculated using not only the data from one LRE but from some superposition of many. Therefore, the gradient minimization of such an error also makes it possible to find not the true current values of the unknown time-varying parameters, but their superposition over time. A more detailed review and analysis of this problem could be found in [15]. Here it should be noted that some integral-based filtering methods, which use the non-stationary aperiodic filters [10] to implement the forgetting of the already irrelevant data about the "outdated" LREs, allow one to identify the piecewise-constant unknown LRE parameters, but only if the PE condition is met. Other methods [16-18], which are related to the concurrent learning [5], theoretically allow to remove erroneous data from the stack even when the PE condition is not met. However, despite the above-stated property, most experiments in [16-18] were conducted when the regressor was PE, which undoubtedly raises questions.

Since the PE condition is rather restrictive, and the identification of the time-varying parameters of LRE is the main motivation to apply the online identification algorithms in practice, the problem of their inaccurate identification with the help of the integral-based filtering methods is actual.

In general, the unknown time-varying parameters of LRE can be of three types: piecewise-constant (switched), piecewise-continuous

Research was financially supported by Russian Foundation for Basic Research (Grant 18-47-310003-r_a).

A. I. Glushchenko is with Stary Oskol technological institute (branch) NUST "MISIS", Stary Oskol, Russia (phone: +79102266946; e-mail: a.glushchenko@sf-misis.ru).

V. A. Petrov is with Stary Oskol technological institute (branch) NUST "MISIS", Stary Oskol, Russia (e-mail: 79040882508@ya.ru).

K. A. Lastochkin is with Stary Oskol technological institute (branch) NUST "MISIS", Stary Oskol, Russia (e-mail: lastconst@yandex.ru).

and continuous time-varying ones. In this paper, the identification of the piecewise-constant unknown parameters of LRE is the only problem of interest.

Earlier in [15], a method of identification of the piecewise-constant parameters, based on the I-DREM procedure and the integral filtering method with exponential forgetting and resetting, has been proposed. Using it, the time point to reset the filter is determined by some external signal, which, for example, considering MRAC problem, is related to the moment of change of the setpoint value. With regard to the above-mentioned time point, three possible cases of jump change of the LRE parameters are singled out. The unknown parameters change their values before the reset time (*case 1*), over the interval of the finite excitation, which is caused by the setpoint value change (*case 2*), after the regressor finite excitation interval is over (*case 3*). The method of filtering with resetting, which has been proposed in [15], allows one to identify the piecewise-constant unknown parameters only in *case 1*. Therefore, in this paper, it is proposed to improve the integral-based filtering with exponential forgetting and resetting so as to extend the results obtained in [15] to *case 2* as well.

The concept of the proposed modification consists in the idea of integration of the filter equations [15] over the finite time intervals, which fully belong to the finite excitation interval, instead of the previously used integration over the infinite time interval. It will be shown that such a change allows one to provide the estimation error to be exponentially bounded over the regressor excitation time range in the case of the piecewise-constant parameters of LRE and to ensure exponential convergence of the estimation error to zero when the LRE parameters are constant after the regressor excitation is over.

The structure of the paper is as follows. The basic definitions and notations used in the paper are presented in Section II; Section III contains a generalized formal problem statement; the proposed modification of the integral-based filter with exponential input decay is in Section IV; Section V is to analyze the stability of the derived estimation law; in Section VI the numerical simulation experiments are conducted.

## II. PRELIMINARIES

The following notations are used throughout the paper. $L_\infty$ is the space of essentially bounded functions, $\|.\|$ is the Euclidean norm of a vector, *floor*(.) is the operation of rounding values to integers, $f(t)$ is a function, which depends on time (for the sake of brevity, the argument $t$ is omitted in some cases). Also, the following definitions will be often used:

**Definition 1:** *The regressor* $\varphi(t) \in L_\infty$ *is finitely exciting* ($\varphi \in$FE) *over the interval* $[t_s; t_s+T]$ *if there exist* $t_s \geq 0$, $T > 0$ *and* $\alpha > 0$ *such that the following inequality holds:*

$$\int_{t_s}^{t_s+T} \varphi(\tau)\varphi^T(\tau)d\tau \geq \alpha I, \qquad (1)$$

*where I is an identity matrix, $\alpha$ is the excitation level.*

**Definition 2:** *The scalar regressors* $\omega_j(t) \in L_\infty$ *are finitely exciting* ($\omega_j \in$FE) *over the same time interval* $[t_s; t_s+T_s]$ *if there exist* $t_s \geq 0$, $T_s > 0$ *and* $\alpha_j > 0$ *such that the following inequalities hold:*

$$\int_{t_s}^{t_s+T_s} \omega_j^2(\tau)d\tau \geq \alpha_j. \qquad (2)$$

*where $\alpha_j$ is an excitation level of the $j^{th}$ regressor.*

The definitions of time interval boundedness (IB), exponential time interval boundedness (IEB), global exponential stability (GES) and exponential ultimate boundedness (EUB) are introduced to be used in the stability analysis. Let $\dot{x}(t) = f(x(t))$ be a nonlinear

dynamical system, which is globally Lipschitz continuous with an equilibrium point at the origin. Then:

**Definition 3:** *The solution of the system $x(t)$ is bounded over a time interval* ($x \in$IB) *if there exists such an interval* $[\tau_1; \tau_2]$ *for* $\kappa > 0$, $\rho > 0$ *and* $R > 0$ *that* $\forall t \in [\tau_1; \tau_2]$ $\|x(t)\| \leq \rho \|x(\tau_1)\| e^{-\kappa(t-\tau_1)} + R$ *for all $x(0)$.*

**Definition 4:** *The solution of the system $x(t)$ is exponentially bounded over a time interval* ($x \in$IEB) *if there exists such an interval* $[\tau_1; \tau_2]$ *for* $\kappa > 0$, $\rho > 0$ *that* $\forall t \in [\tau_1; \tau_2]$ *that* $\|x(t)\| \leq \rho \|x(\tau_1)\| e^{-\kappa(t-\tau_1)}$ *for all $x(0)$.*

**Definition 5:** *The equilibrium point of the system is globally exponentially stable* ($x \in$GES) *if there exists such time moment T for* $\kappa > 0$ *and* $\rho > 0$ *such that* $\|x(t)\| \leq \rho \|x(0)\| e^{-\kappa t}$ *for* $t \geq T$ *and all $x(0)$.*

**Definition 6:** *The solution $x(t)$ is exponentially ultimately bounded* ($x \in$EUB) *with the uniform ultimate bound R if there exists such time moment T for* $\kappa > 0$, $\rho > 0$ *and* $R > 0$ *that* $\|x\| \leq \rho \|x(0)\| e^{-\kappa t} + R$ *for* $t \geq T$ *and all $x(0)$.*

## III. PROBLEM STATEMENT

Let the linear regression equation with the scalar regressor be considered. It could be derived from any initial vector LRE by application of the DREM procedure [12]:

$$y(t) = \omega(t)\Theta, \qquad (3)$$

where $y \in R^n$ is a measurable function, $\omega \in R$ is a measurable scalar regressor, $\Theta \in R^n$ is a vector of unknown parameters.

It is assumed that the regressor $\omega$ is finitely exciting (1) over some interval $[t_r^+; t_e]$. The jump change of the unknown parameters $\Theta$ values is matched with the excitation interval as follows:

**Definition 7.** *The unknown parameters $\Theta$ $\forall t \geq t_r^+$ are such as:*

$$\dot{\Theta} = 0; \ \Theta = \Theta_0 + \sum_i \theta_i h(t - t_i), \qquad (4)$$

*where $h(t - t_i)$ is a Heaviside step function at the time point $t_i$, $t_i < t_e$ are time points of the unknown parameters $\Theta$ jump change.*

Thus, the class of the unknown piecewise-constant LRE parameters is limited to the vector $\Theta$ elements, which change their values stepwise over the excitation interval, but are constant afterwards. It is required to derive the parameter $\Theta$ estimation technique that ensures that the following requirements is satisfied if $\omega \in$FE:

$$\forall t \in [t_r^+; t_e] : \exists [\tau_1; \tau_2], \ \forall t \in [\tau_1; \tau_2] \ \tilde{\Theta} \in \text{IEB} \qquad (5)$$

$$\forall t \geq t_e : \tilde{\Theta} \in \text{GES},$$

where $\tilde{\Theta} = \hat{\Theta} - \Theta$ is the parameter estimation error.

## IV. MAIN RESULT

The proposed solution to the stated problem of the piecewise-constant LRE parameters estimation includes two regressor preprocessing procedures – normalization of the regressor excitation and the integral-based filtering over a finite time interval. In the following subsections they will be described in more detail.

### A. Excitation Normalization

Let the procedure of the regressor excitation normalization [19] be applied to the regression (3). For this purpose, let the scalar regressor $\omega$ be presented in a numerical notation format:

$$\omega(t) = \text{sgn}(\omega)10^\eta$$

$$\text{sgn}(\omega) = \begin{cases} 1 \text{ if } \omega \geq 0 \\ -1 \text{ otherwise} \end{cases}; \ \eta = \begin{cases} \log_{10}(|\omega|) \text{ if } |\omega| \neq 0 \approx 10^{-\infty} \\ -\infty \text{ otherwise} \end{cases} \qquad (6)$$

Using the definition (6) and in accordance with [19], let the normalization function for $\omega(t)$ be introduced:

$$f(\omega) = \text{sgn}(\omega)10^{-sat(\eta)},$$

$$\text{sat}(\eta) = \begin{cases} \eta_{\min} \ \text{if} \ \eta \le \eta_{\min} \\ \eta \ \text{otherwise} \end{cases}. \quad (7)$$

Let the regression (1) be multiplied by the function (7):

$$Y(t) = y(t)f(\omega) = \omega(t)f(\omega)\Theta = \varphi(t)\Theta$$

$$\varphi(t) = \begin{cases} 10^{\eta-\eta_{\min}}, \ \text{if} \ \eta \le \eta_{\min} \\ 1, \ \text{otherwise} \end{cases} \quad (8)$$

Let the propositions, which are true for the regressor $\varphi(t)$, be copied from [19], where their proof could be found.

**Proposition 1.** *The normalized regressor* $\varphi \in [0; 1]$.

**Proposition 2.** *If* $\omega_j \in$ FE *over the excitation interval* $\left[t_r^+; t_e\right]$, *then for the normalized regressor* $\varphi$ *over the interval* $\left[t_r^+; t_e\right]$:

*1) when* $\eta \le \eta_{\min}$, *the following inequality holds:*

$$10^{-2\eta_{\min}} \alpha_j \le \int_{t_r^+}^{t_e} \varphi^2(\tau) d\tau \le t_e - t_r^+. \quad (9)$$

*2) when* $\eta > \eta_{\min}$, *the following inequality holds:*

$$0 \le \Delta \le \int_{t_r^+}^{t_e} \varphi^2(\tau) d\tau \le t_e - t_r^+, \quad (10)$$

*where* $\Delta$ *is the same value for the regressors* $\omega_j$.

**Proposition 3.** *Let* $\omega_j \in$ FE *and there exists such a time point* $T_j \in \left(t_r^+; t_e\right)$ *over the excitation interval* $\left[t_r^+; t_e\right]$ *that* $\forall t \in \left[t_r^+; T_j\right)$ $\eta > \eta_{\min}$ *holds, while* $\forall t \in \left[T_j; t_e\right]$ *the inequality* $\eta \le \eta_{\min}$ *holds. Then the following is true for the normalized regressor* $\varphi$:

$$0 \le \Delta_{\min} \le \int_{t_r^+}^{t_e} \varphi^2(\tau) d\tau \le T, \quad (11)$$

*where* $\Delta_{\min} \le \min_{j \ge 0}\left\{T_j - t_r^+\right\} \le T_j - t_r^+$ *is the same value for all* $\omega_j$.

*The proofs of Propositions 1-3 is presented in [19].*

The following assumption is made about the parameter $\eta_{\min}$ for the sake of the following analysis:

**Assumption 1.** *The parameter* $\eta_{\min}$ *is chosen so as Proposition 3 holds over the excitation interval* $\left[t_r^+; t_e\right]$.

### B. Filtration over a time window

Let the regression with normalized regressor excitation (8) be filtered as follows:

$$\Omega(t_k) = \Omega(T^+) = \Upsilon(t_k) = \Upsilon(T^+) = 0$$

$$\Omega(t) = \begin{cases} \int_{t_k}^{t} \exp\left(-\int_0^{\tau} \sigma d\tau_1\right) \varphi^2(\tau) d\tau, \ \text{if} \ t < T^+; \\ \int_{T^+}^{t} \exp\left(-\int_0^{\tau} \sigma d\tau_1\right) \varphi^2(\tau) d\tau, \ \text{if} \ t \ge T^+ \end{cases}; \quad (12)$$

$$\Upsilon(t) = \begin{cases} \int_{t_k}^{t} \exp\left(-\int_0^{\tau} \sigma d\tau_1\right) Y(\tau)\varphi(\tau) d\tau, \ \text{if} \ t < T^+ \\ \int_{T^+}^{t} \exp\left(-\int_0^{\tau} \sigma d\tau_1\right) Y(\tau)\varphi(\tau) d\tau, \ \text{if} \ t \ge T^+ \end{cases};$$

where $0 < T < t_e - t_r^+$ is the filter window width, $0 < T^+ < t_e$ is the time moment, when the filtering with the help of the time windows is stopped, and $t_k = T \cdot floor(t/T)$, $\sigma > 0$.

Let the properties of the obtained regressor $\Omega(t)$ be described over the intervals $\left[t_k; t_{k+1}\right]$ and $\left[T^+; t\right]$ when $\omega_j \in$ FE. For this purpose, the following assumption of the continuity of the regressor $\omega_j$ excitation over the interval $\left[t_r^+; t_e\right]$ is introduced:

**Assumption 2.** *If the regressor* $\omega_j \in$ FE *over the interval*

$\left[t_r^+; t_e\right]$, *then* $\forall t \in \left[t_r^+; t_e\right] \ \exists T_s > 0$ *such that* $\omega_j \in$ FE *over the interval* $\left[t; t + T_s\right] \subset \left[t_r^+; t_e\right]$.

Assumption 2 holds, particularly, if the regressor $\omega_j$ is continuous and does not converge to zero over finite time during the excitation interval $\left[t_r^+; t_e\right]$. This is usually true in practice.

**Proposition 4.** *If* $\omega_j \in$ FE *over the interval* $\left[t_r^+; t_e\right]$ *and the assumptions 1 and 2 holds, then the following is true for* $\Omega(t)$:

*1)* $\forall t \in \left[t_r^+; t_e\right] \ \Omega(t) \ge 0$;

*2)* $\exists T_{0k} \in [t_k; t_{k+1}].$ $\forall t \in [T_{0k}; t_{k+1}]$ $0 < \Omega(T_{0k}) \le \Omega(t) \le T$.

*Proof.*

According to (12), $sign\left(\exp\left(-\int_0^t \sigma d\tau_1\right)\varphi^2(\tau)\right) \ge 0$. Then, $\dot{\Omega}(t) \ge 0$ and $\Omega(t) \ge 0$ $\forall t \in \left[t_r^+; t_e\right]$.

To prove the second part of the proposition, let the estimate of the exponentially decaying multiplier from the integrand of the regressor definition $\Omega(t)$ be written:

$$\forall t \in [t_k; t_{k+1}]: \exp(-\sigma T) \le \exp\left(-\int_0^t \sigma d\tau_1\right) \le 1 \quad (13)$$

According to Assumption 2, $\omega_j \in$ FE over the interval $[t_k; t_{k+1}]$. Then, using the propositions 2-3, three situations are possible over the interval $[t_k; t_{k+1}]$:

*1)* $\forall t \in [t_k; t_{k+1}] \ \eta \le \eta_{\min}$;

*2)* $\forall t \in [t_k; t_{k+1}] \ \eta > \eta_{\min}$;

*3)* $\forall t \in [t_k; T_j) \ \eta > \eta_{\min}$, whereas $\forall t \in \left[T_j; t_{k+1}\right] \ \eta \le \eta_{\min}$.

Combining the left-hand sides of equations (9)-(11), which are written for the regressor $\omega_j$ over the interval $[t_k; t_{k+1}]$, it is obtained that:

$$0 < \underbrace{\min\left\{\min_{j \ge 0}\left\{T_j - t_k\right\}; \Delta_k; \min_{j \ge 0, k \ge 0}\left\{10^{-2\eta_{\min}}\alpha_{jk}\right\}\right\}}_{\Delta_1} \le \int_{t_k}^{t_{k+1}} \varphi^2(\tau) d\tau \le T, \quad (14)$$

where $\Delta_1$ is the regressor $\varphi$ excitation level over the interval $[t_k; t_{k+1}]$, $0 < \Delta_k \le t_{k+1} - t_k$ is the same value for the regressors $\omega_j$, $\alpha_{jk}$ is the excitation level of the $j^{th}$ regressor over the interval $[t_k; t_{k+1}]$.

Taking into account the estimates (13) and (14) and using the mean value theorem, the following estimate of $\Omega(t_{k+1})$ is obtained:

$$0 < \exp(-\sigma T)\Delta_1 \le \int_{t_k}^{t_{k+1}} \exp\left(-\int_0^t \sigma d\tau_1\right)\varphi^2(\tau) d\tau \le T \quad (15)$$

As $\Omega(t) \ge 0$ $\forall t \in [t_k; t_{k+1}]$ and $\Omega(t_{k+1}) \ge \exp(-\sigma T)\Delta_1$, then $\exists T_{0k} \in [t_k; t_{k+1}]$ such as $\Omega(T_{0k}) = \exp(-\sigma T)\Delta_1$ and the following holds $\forall t \in [T_{0k}; t_{k+1}]$:

$$0 < \Omega(T_{0k}) \le \Omega(t) \le T, \quad (16)$$

as was to be proved in the second part of Proposition 4.

**Proposition 5.** *If* $t \ge T^+$, $\omega_j \in$ FE *over the interval* $\left[t_r^+; t_e\right]$ *and Assumptions 1 and 2 hold, then* $\forall t \ge t_e$ $0 < \Omega_{LB} \le \Omega(t) \le \Omega_{UB}$.

*Proof.*

Considering $\varphi \in [0; 1]$, to prove this proposition, first of all, the upper bound estimate of the regressor $\Omega(t)$ $\forall t \ge T^+$ is obtained:

$$\Omega(t) = \int_{T^+}^{t} \exp\left(-\int_0^t \sigma d\tau_1\right)\varphi^2(\tau) d\tau \le \int_{T^+}^{t} \exp\left(-\int_0^t \sigma d\tau_1\right) d\tau \le \frac{1}{\sigma} = \Omega_{UB}. \quad (17)$$

Then the regressor $\Omega(t)$ $\forall t \ge t_e$ is redefined as:

$$\Omega(t) = \int_{T^+}^{t_e} \exp\left(-\int_0^t \sigma d\tau_1\right)\varphi^2(\tau) d\tau + \int_{t_e}^{t} \exp\left(-\int_0^t \sigma d\tau_1\right)\varphi^2(\tau) d\tau. \quad (18)$$

Let the first integral of the sum (18) be considered. According to Assumption 2, $\omega_j \in$ FE over the interval $\left[T^+; t_e\right]$. Then, applying Propositions 2-3, three situations are possible:

1) $\forall t \in \left[ T^{+}; \ t_{e} \right] \ \eta \le \eta_{min}$ ;

2) $\forall t \in \left[ T^{+}; \ t_{e} \right] \ \eta > \eta_{min}$ ;

3) $\forall t \in \left[ T^{+}; \ T_{j} \right) \ \eta > \eta_{min}$, whereas $\forall t \in \left[ T_{j}; \ t_{e} \right] \ \eta \le \eta_{min}$.

Combining the left-hand sides of equations (9)-(11), which are written for the regressor $\varphi$ over the interval $\left[ T^{+}; \ t_{e} \right]$, it is obtained:

$$0 < \underbrace{\min \left\{ \min_{j \ge 0} \{ T_{j} - T^{+} \}; \Delta_{0k}; \min_{j \ge 0; \ k \ge 0} \{ 10^{-2\eta_{min}} \alpha_{0jk} \} \right\}}_{\Delta_{2}} \le \int_{T^{+}}^{t_{e}} \varphi^{2}(\tau) d\tau. \ (19)$$

where $\Delta_{2}$ is the regressor $\varphi$ excitation level over the interval $\left[ T^{+}; \ t_{e} \right]$, $0 < \Delta_{0k} \le t_{e} - T^{+}$ is the same value for the regressors $\omega_{j}$, $\alpha_{0jk}$ is the excitation level of the $j^{\text{th}}$ regressor over the interval $\left[ T^{+}; \ t_{e} \right]$.

Considering (19), similarly to (15), the lower-bound estimate of the first integral of the sum (18) is written as:

$$0 < \underbrace{\Delta_{2} \exp \left( -\sigma \left( t_{e} - T^{+} \right) \right)}_{\Omega_{LB}} \le \int_{T^{+}}^{t_{e}} \exp \left( -\int_{0}^{\tau} \sigma d\tau_{1} \right) \varphi^{2}(\tau) d\tau. \ (20)$$

Combining (17) and (20), the following inequality is obtained, which holds $\forall t \ge t_{e}$ :

$$0 < \Omega_{LB} \le \Omega(t) \le \Omega_{UB}, \ (21)$$

which completes the proof of Proposition 5.

Then some important features of the regressor $\Omega(t)$ will be shown. They are guaranteed if $T^{+}$ value is chosen according to the condition $T^{+} \le \min_{j \ge 0} \{ T_{j} \}$. So, such an assumption about $T^{+}$ is made, and the properties of the regressor $\Omega(t)$ are analyzed again.

**Assumption 3.** $T^{+}$ is chosen so as $T^{+} \le \min_{j \ge 0} \{ T_{j} \}$.

**Proposition 6.** If $t < T^{+}$, $\omega_{j} \in$ FE *over the interval* $\left[ t_{e}^{+}; \ t_{e} \right]$, *the assumptions 1-3 hold, then* $\exists \overline{T}_{0k} \in \left[ t_{k}; \ t_{k+1} \right]$ *such that* $\forall t \in \left[ \overline{T}_{0k}; \ t_{k+1} \right] \ 0 < \Omega(t) \le T$, *and* $\Omega \left( \overline{T}_{0k} \right)$ *is the same value* $\forall k$ *for the regressors* $\omega_{j}$.

*Proof.*

To prove the proposition, let it be considered that $t < T^{+} \ \forall t \in \left[ t_{k}; \ t_{k+1} \right]$ and $T^{+}$ is chosen as $T^{+} \le \min_{j \ge 0} \{ T_{j} \}$.

Then, according to Assumption 3, the following holds: $\eta > \eta_{min}$ $\forall t \in \left[ t_{k}; \ t_{k+1} \right]$. It means that, considering (6), $\varphi = 1 \ \forall t \in \left[ t_{k}; \ t_{k+1} \right]$. As a result, in a similar manner to the second part of Proposition 2, it is obtained that:

$$0 < \Delta_{k} \le \int_{t_{k}}^{t_{k+1}} \varphi^{2}(\tau) d\tau \le T, \ (22)$$

where $\Delta_{k} \le t_{k+1} - t_{k}$ is the same value for the regressors $\omega_{j}$ (the normalized value of the regressor $\varphi$ excitation level over $\left[ t_{k}; \ t_{k+1} \right]$).

Taking into consideration the estimates (13) and (22) and applying the mean value theorem, the bounds of $\Omega(t_{k+1})$ are obtained:

$$0 < \exp \left( -\sigma T \right) \Delta_{k} \le \int_{t_{k}}^{t_{k+1}} \exp \left( -\int_{0}^{\tau} \sigma d\tau_{1} \right) \varphi^{2}(\tau) d\tau \le T. \ (23)$$

As $\Omega(t) \ge 0 \ \forall t \in \left[ t_{k}; \ t_{k+1} \right]$ and $\Omega(t_{k+1}) \ge \exp \left( -\sigma T \right) \Delta_{k}$, then $\exists \overline{T}_{0k} \in \left[ t_{k}; \ t_{k+1} \right]$ such that $\Omega \left( \overline{T}_{0k} \right) = \exp \left( -\sigma T \right) \Delta_{k}$, and $\forall t \in \left[ \overline{T}_{0k}; \ t_{k+1} \right]$ it holds that:

$$0 < \Omega \left( \overline{T}_{0k} \right) \le \Omega(t) \le T. \ (24)$$

As, when $T^{+} \le \min_{j \ge 0} \{ T_{j} \}$, $\varphi = 1$ independently from $k$ and $\omega_{j}$ values, then $\Omega \left( \overline{T}_{0k} \right)$ is the same value $\forall k$ for the regressors $\omega_{j}$. This completes the proof.

**Proposition 7.** If $\omega_{j} \in$ FE *over the interval* $\left[ t_{e}^{+}; \ t_{e} \right]$, *Assumptions 1-3 are met, then* $\forall t \ge t_{e} \ 0 < \overline{\Omega}_{LB} \le \Omega(t) \le \Omega_{UB}$, $\overline{\Omega}_{LB}$ *is the same value* $\forall k$ *for the regressors* $\omega_{j}$.

*Proof.*

To prove the proposition, let the first integral of the sum (18) be considered. According to Assumption 2, $\omega_{j} \in$ FE over the interval $\left[ T^{+}; \ t_{e} \right]$ and, because $T^{+}$ is chosen under the condition $T^{+} \le \min_{j \ge 0} \{ T_{j} \}, \forall t \ge T^{+}$, we have $\eta \ge \eta_{min}$ in the most conservative case. Then, in accordance with the first part of Proposition 2, the following inequality is written:

$$10^{-2\eta_{min}} \alpha_{min} \le \int_{T^{+}}^{t_{e}} \varphi^{2}(\tau) d\tau \le t_{e} - T^{+}. \ (25)$$

where $\alpha_{min} = \min_{j \ge 0} \{ \alpha_{0j} \}$ is the same value for the regressors $\omega_{j}$, $\alpha_{0j}$ is the excitation level of the $j^{\text{th}}$ regressor over the interval $\left[ T^{+}; \ t_{e} \right]$.

Taking into account (25), in a similar way to (15), the lower bound of the first integral of the sum (18) is written as:

$$0 < \underbrace{10^{-2\eta_{min}} \alpha_{min} \exp \left( -\sigma \left( t_{e} - T^{+} \right) \right)}_{\overline{\Omega}_{LB}} \le \int_{T^{+}}^{t_{e}} \exp \left( -\int_{0}^{\tau} \sigma d\tau_{1} \right) \varphi^{2}(\tau) d\tau. \ (26)$$

where $\overline{\Omega}_{LB}$ is the same value for the regressors $\omega_{j}$.

Combining (17) and (26), the following inequality holds $\forall t \ge t_{e}$ for the regressor $\Omega(t)$:

$$0 < \overline{\Omega}_{LB} \le \Omega(t) \le \Omega_{UB}, \ (27)$$

which completes the proof of Proposition 7.

The gradient-based estimation law, which uses the regressor $\Omega(t)$ and function $\Upsilon(t)$, is written as:

$$\dot{\hat{\Theta}} = \dot{\tilde{\Theta}} = -\gamma \Omega \left( \Omega \hat{\Theta} - \Upsilon \right). \ (28)$$

where $\gamma$ is the estimation law adaptation rate.

**Remark 1.** *Propositions 4 and 5 (as well as 6 and 7) are equiform. They show that, considering the time intervals* $\left[ t_{k}; \ t_{k+1} \right]$ *and* $\left[ T^{+}; \ t \right]$, *the regressor* $\Omega(t)$ *is bounded from below by a non-zero value on and after a certain time point. As far as the value of $T^{+}$ is chosen in accordance with Assumption 3, the application of the regressor excitation normalization procedure* (6)-(7) *allows one to obtain the integral-based regressor* $\Omega(t)$ *with the same lower bound for all initial regressors* $\omega_{j}$. *As it will be shown in the next section, this fact gives the opportunity to use the same constant value of $\gamma$ for various initial regressors of the estimation law* (28).

## V. STABILITY ANALYSIS

In this section of the paper, the stability of the identification law (28) will be analyzed when the piecewise-constant parameters (4) of the regression (3) are estimated.

Based on the problem statement, three situations are possible over the interval $\left[ t_{r}^{+}; \ t_{e} \right]$: 1) $t_{i} \in \left[ t_{k}; \ t_{k+1} \right]$, so the time point of the regression parameters jump change belongs to the range of the integration over the interval $\left[ t_{r}^{+}; \ T^{+} \right]$; 2) $t_{i} \in \left[ T^{+}; \ t_{e} \right]$, so the time point of the regression parameters jump change belongs to the range of integration over the interval $\left[ T^{+}; \ t_{e} \right]$; 3) $\forall t \ge t_{r}^{+} \ \Theta = \Theta_{0}$, so the regression parameters do not change their values over the excitation interval. The following theorems are to consider these cases in more detail.

**Theorem 1.** *Let* $t \in \left[ t_{r}^{+}; \ T^{+} \right]$, *all requirements of Propositions 6 and 7 be met, the time point of the jump change of the regression parameters* $t_{i} \in \left[ t_{k}; \ t_{k+1} \right]$, *and there exists the time range* $\left[ t_{k+1}; \ t_{k+2} \right]$ *such that* $t_{k+2} \le T^{+}$, $t_{i+1} \notin \left[ t_{k+1}; \ t_{k+2} \right]$, *then:*

1) $\forall t \in \left[ \overline{T}_{0k}; \ t_{k+1} \right] \ \dot{\tilde{\Theta}}(t) \in$ IB ;

2) $\forall t \in \left[ \overline{T}_{0(k+1)}; \ t_{k+2} \right] \ \dot{\tilde{\Theta}}(t) \in$ IEB .

*The proof is postponed to subsection A of Appendix.*

**Theorem 2.** Let all requirements of Propositions 6 and 7 be met, and $t_i \in \left\lfloor T^+ ; t_e \right\rfloor$, then:

1) $\forall t \in \left\lfloor \overline{T}_{0k} ; t_{k+1} \right\rfloor \tilde{\Theta}(t) \in \text{IEB}$;

2) $\forall t \geq t_e \ \tilde{\Theta}(t) \in \text{EUB}$.

The proof is in subsection B of Appendix.

**Theorem 3.** Let $\forall t \geq t_r^* \ \Theta = \Theta_0$ and the requirements of Propositions 6 and 7 be met, then:

1) $\forall t \in \left\lfloor \overline{T}_{0k} ; t_{k+1} \right\rfloor \tilde{\Theta}(t) \in \text{IEB}$;

2) $\forall t \geq t_e \ \tilde{\Theta}(t) \in \text{GES}$.

The proof is in subsection C of Appendix.

**Remark 2.** Proposition 3 is expected to be met in Theorems 1-3. However, as the pairs of propositions 4 and 5, 6 and 7 are equiform respectively, this proposition does not lead to loss of the generality of the results obtained in the theorems (from the point of view of the stability of the parameter error) and allows one to use any $0 < T^+ < t_e$. If the value of $T^+$ is chosen under the condition $T^+ \leq \min_{j \geq 0} \{ T_j \}$, then the following is obtained: 1) boundedness of $\tilde{\Theta}$ from above by the same number for regressors $\omega_j$; 2) exponential convergence of $\tilde{\Theta}$ to zero or a set with the same minimum convergence rate for all $\omega_j$.

The following corollaries can be derived from these theorems.

**Corollary 1.** If $t_i \in [t_k; t_{k+1}]$ and there exists the time interval $[t_{k+1}; t_{k+2}]$ such that $t_{k+2} \leq T^+$, $t_{i+1} \in [t_{k+1}; t_{k+2}]$, then $\exists \left\lfloor \overline{T}_{0(k+1)} ; t_{k+2} \right\rfloor$, $\forall t \in \left\lfloor \overline{T}_{0(k+1)} ; t_{k+2} \right\rfloor \tilde{\Theta}(t) \in \text{IEB}$.

**Corollary 2.** If $\nexists t_i \in \left\lfloor T^+ ; t_e \right\rfloor$, then $\forall t \geq t_e \ \tilde{\Theta} \in \text{GES}$.

In other words, the first requirement of the objective (5) is met if, considering the time interval $\left\lfloor t_r^* ; t_e \right\rfloor$, the regression parameters have changed their values over $[t_k; t_{k+1}]$, and there exists a time range $[t_{k+1}; t_{k+2}]$, during which the regression parameters are constant. And the second requirement is met if the regression parameters do not change their values over the interval $\left\lfloor T^+ ; t_e \right\rfloor$. Obviously, it is possible to improve the probability that the first requirement is met by reduction of the value of $T$. Thereby we will increase the total number of time windows to apply the filter (12) and, consequently, the sensitivity of the filter (12) to changes of the regression parameters. It is possible to improve the probability that the second requirement is met if $T^+ \to t_e$. However, the conditions $T^+ \to t_e$ and $T^+ \leq \min_{j \geq 0} \{ T_j \}$ are often mutually exclusive.

## VI. NUMERICAL EXAMPLE

In this section, the numerical modelling is used to demonstrate the above-stated properties of the proposed estimation law. If there are no disturbances, $\tilde{\Theta} \in \text{GES}$ and $\tilde{\Theta} \in \text{IEB}$. And these are sufficient and necessary to provide $\tilde{\Theta} \in \text{EUB}$ and $\tilde{\Theta} \in \text{IB}$ in case of perturbations respectively. So, the numerical experiments have been conducted for both cases.

### A. Noise-free case.

The regressor $\omega$ is an output of the aperiodic filter:

$$\dot{\omega}_j(t) = -\omega_j(t) + u_j, \ \omega_j(0) = 0 \quad (29)$$

Three regressors are considered. They are output signals of (29) when $u_j$ is chosen as:

$$u_1 = e^{-t}; \ u_2 = 10e^{-t}; \ u_3 = 100e^{-t}. \quad (30)$$

As far as the applied problems are concerned, the regressors, which are obtained by feeding exponentially decaying signals (30) to the filter (29) input, are most similar, considering their shape, to the scalar regressors, which are typically formed by the DREM procedure. In addition, the regressor, which is obtained from the output of (29), meets the requirements of Assumption 2.

The values of when and how to change the unknown parameters (4) of the regression (3) were defined according to the expression:

$$\Theta_0 = 1; \ (t_i, \theta_i) = \begin{cases} t_1 = 0.25; \theta_1 = 0.5 \\ t_2 = 1.05; \theta_2 = 0.5 \\ t_3 = 1.45; \theta_3 = -0.5 \\ t_4 = 2.25; \theta_3 = 0.5 \\ t_5 = 2.75; \theta_4 = -0.5 \end{cases} \quad (31)$$

The parameters of the normalization function (7), the filter (12), and the estimation law (28) were defined as follows:

$$\eta_{min} = -1; \ T = 0.5; \ T^+ = 3; \ \gamma = 10^6; \ \hat{\Theta}(0) = 0; \ \sigma = \frac{s}{2T} \quad (32)$$

Before presenting the results of the experiment, we tried to predict them using Theorems 1-3. The results of such prediction are presented as Table I.

TABLE I
THEORETICAL PREDICTION OF SIMULATION RESULTS

| Time Interval | Type of boundedness | Time Interval | Type of boundedness |
|---|---|---|---|
| $t \in [0; 0.25]$ | $\tilde{\Theta} \in \text{IEB}$ | $t \in [1.5; 2]$ | $\tilde{\Theta} \in \text{IEB}$ |
| $t \in [0.25; 0.5]$ | $\tilde{\Theta} \in \text{IB}$ | $t \in [2; 2.25]$ | $\tilde{\Theta} \in \text{IEB}$ |
| $t \in [0.5; 1]$ | $\tilde{\Theta} \in \text{IEB}$ | $t \in [2.25; 2.5]$ | $\tilde{\Theta} \in \text{IB}$ |
| $t \in [1; 1.05]$ | $\tilde{\Theta} \in \text{IEB}$ | $t \in [2.5; 2.75]$ | $\tilde{\Theta} \in \text{IEB}$ |
| $t \in [1.05; 1.45]$ | $\tilde{\Theta} \in \text{IB}$ | $t \in [2.75; 3]$ | $\tilde{\Theta} \in \text{IB}$ |
| $t \in [1.45; 1.5]$ | $\tilde{\Theta} \in \text{IB}$ | $t \geq 3$ | $\tilde{\Theta} \in \text{GES}$ |

Having made the prediction, the numerical simulation was conducted. Figure 1 shows transients of regressors $\omega_j$, $\varphi_j$ and $\Omega_j$.

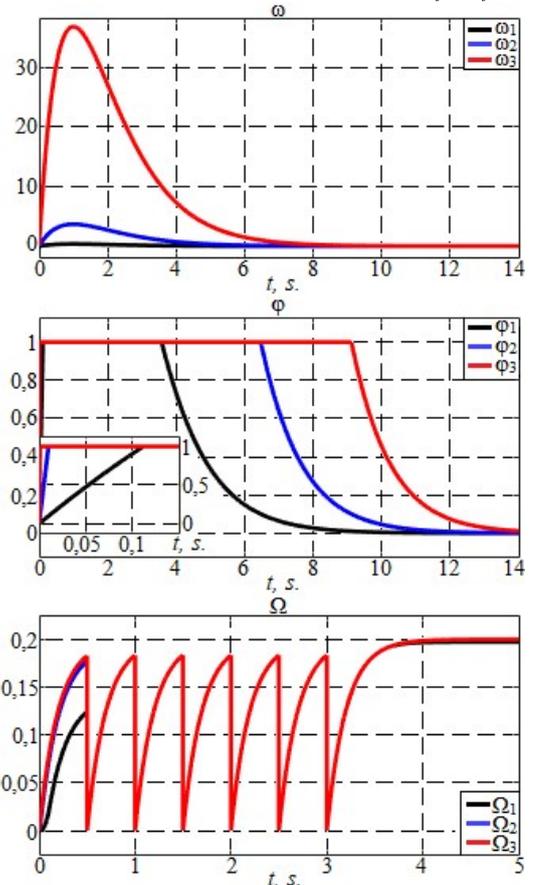

Fig. 1. Transients curves of regressors $\omega_j$, $\varphi_j$ and $\Omega_j$.

As it follows from the transient curves of the regressors $\varphi_j$ and $\Omega_j$, Assumption 1 was met from the time point $t \approx 0,112$, Assumption 3 was met as $T^+ \leq 3,47 \leq 6,47 \leq 9,1$. As a result, all needed assumptions were held.

Figure 2 shows the transients of the unknown parameters estimates formed by the law (28) for the regressors $\Omega_j$. Comparison of the transients presented in Fig. 2 with the predicted data from Table I allowed us to fully validate the theoretical conclusions of Theorems 1-3. As follows from Fig. 2, starting from the time point $t = 0,5$ s, the parameter error of the unknown parameter estimates for the various regressors $\omega_j$: 1) was bounded from above by the same value; 2) converged at the same rate to zero (on account of the normalization procedure of the regressor excitation (8)).

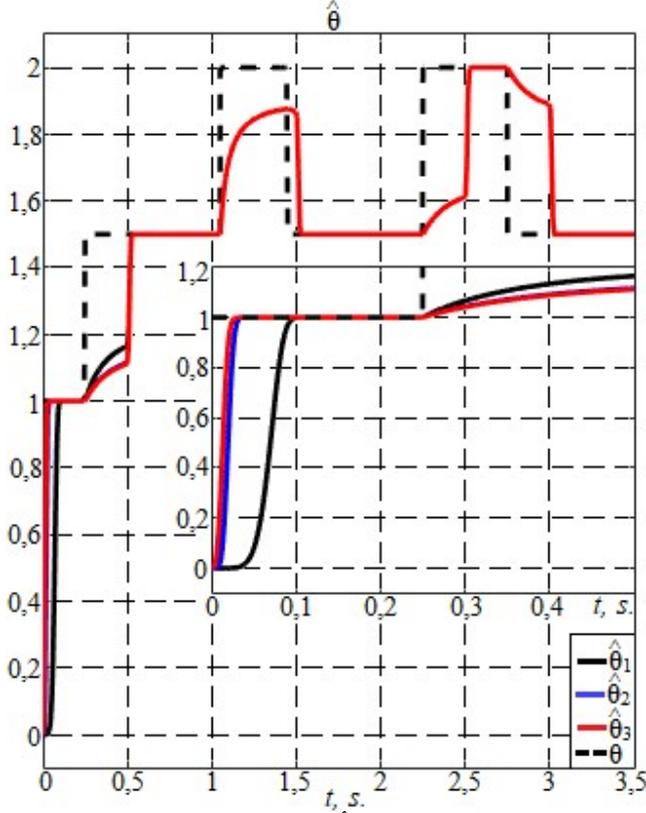

Fig. 2. Transient curves of estimates of $\hat{\theta}_j$ when $T = 0,5$ s.

The time point $t = 0.5$ s was caused by: 1) the fact that all the conditions of Theorems 1-3 were met after $t \approx 0.112$ s, 2) the width of the filtration time window $T = 0.5$ s, so the filter (12) was reset for the first time at the time point $t = 0.5$ s.

Section V was completed with the conclusion that it was possible to improve the quality of the unknown piecewise-constant parameters estimation by a reduction of the parameter $T$ value. This statement was verified experimentally. $T = 0.05$ s, and values of all other filter (12) and law (28) parameters coincided with the values given in (32). Figure 3 shows transient curves of the estimates of the unknown parameters, which were obtained with the law (28) and $T = 0.05$ s.

As follows from the transients shown in Fig. 3, indeed, as the parameter $T$ was reduced, the number of intervals, when $\tilde{\Theta} \in \text{IEB}$, increased. Hence, the quality of the estimation of the unknown parameters improved.

It was noted in Remarks 1, 2 that conclusions, which were drawn in Theorems 1-3 and Corollaries 1, 2 regarding the stability of $\tilde{\Theta}$, were also correct even when Assumption 3 was not met. This was shown by the following example. The equations of $u_j$ were equal:

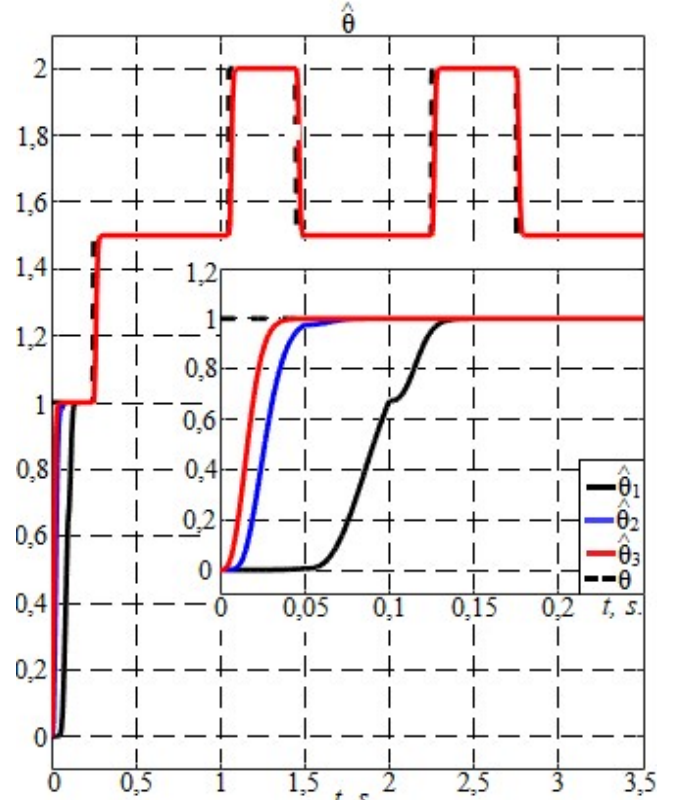

Fig. 3. Transient curves of $\hat{\theta}_j$ when $T = 0.05$ s.

$$u_1 = e^{-5t}; \ u_2 = 10e^{-5t}; \ u_3 = 100e^{-5t}; \qquad (33)$$

All parameters of the normalization procedure (7), (8), interval-based integral filtering (12) and estimation law (28) were chosen according to (32). Figure 4 presents the transient curves of the regressors $\varphi_j$.

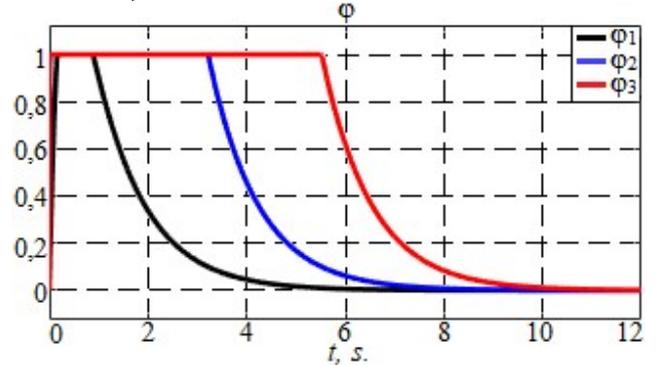

Fig. 4. Transient curves of regressors $\varphi_j$.

Considering the transient curves of $\varphi_j$, the requirement $T^+ \leq 0,88 \leq 3,21 \leq 5,52$ was not met as $3 \geq 0,88$. Hence, in accordance with Remarks 1 and 2, the convergence rate and the value of the parameter error were individual for each of the regressors $\Omega_1$ and $\Omega_2, \Omega_3$.

Figure 5 depicts the transients of the estimates of the unknown parameters obtained with the law (28) for the regressors $\Omega_j$. Comparison of the transient curves from Fig. 5 with the predictions from Table I allowed us to confirm Remarks 1 and 2.

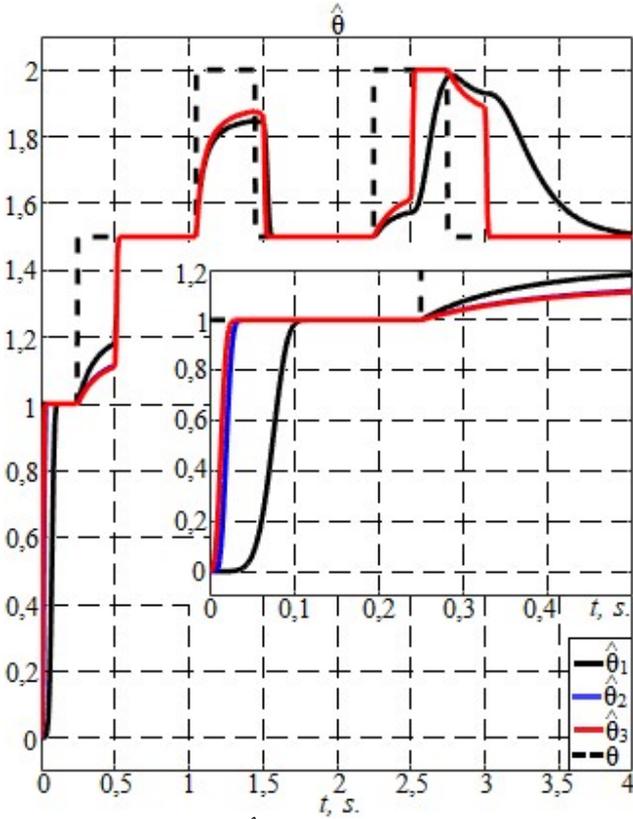

Fig. 5. Transient curves of $\hat{\theta}_j$ when convergence rate is individual for each regressor.

Indeed, if Assumption 3 was not met, the conclusions drawn in Theorems 1-3 and Corollaries 1, 2 concerning stability $\tilde{\Theta}$ were correct, but the convergence rates and error values for the regressors $\Omega_1$ and $\Omega_2, \Omega_3$ did not coincide to each other respectively.

### B. Noise-contaminated case

The LRE (3) with perturbed measurements of the function $y(t)$ and the regressor $\omega(t)$ was considered:

$$y_j(t) = \omega_j(t)\Theta + w_j \qquad (34)$$

Three LRE with regressors $\omega_j$ ($j = 1,2,3$) were considered. Each $\omega_j$ was the output of the filter (29) when the respective $u_j$ from (30) was fed to its input.

The measurable signal to be processed with the help of (7), (8) and (12) was the regressor $\hat{\omega}_j$, which was the sum of the real regressor $\omega_j$ value and the noise component $W_l$:

$$\hat{\omega}_j = \omega_j + W_1, \qquad (35)$$

where $W_l$ was the white noise function with the parameters, which values are shown in Table II.

The noised components of $y_j(t)$ measurement were defined as follows:

$$w_1 = W_2; \ w_2 = 10W_2; \ w_3 = 100W_2. \qquad (35)$$

where $W_2$ was the white noise function with the parameters, which values are shown in Table II.

TABLE II
DISTURBANCE PARAMETERS

| Disturbance | Noise Power | Sample Time | Seed |
|---|---|---|---|
| $W_l$ | $10^{-8}$ | $10^{-4}$ | 23341 |
| $W_2$ | $10^{-7}$ | $10^{-4}$ | 33341 |

The aim of the experiment was to show that the proposed estimation law could function under conditions of the noised data.

Figure 6 presents the transient curves of the regressors $\varphi_j$ and the functions $y_j$. The transient curves of the regressors $\varphi_j$ demonstrated the efficiency of the normalization procedure (7), (8) under the condition of the noised regressor measurement. Transients of functions (34) gave information about the magnitude of perturbations. Figure 7 presents transient curves of the estimates of the unknown parameters.

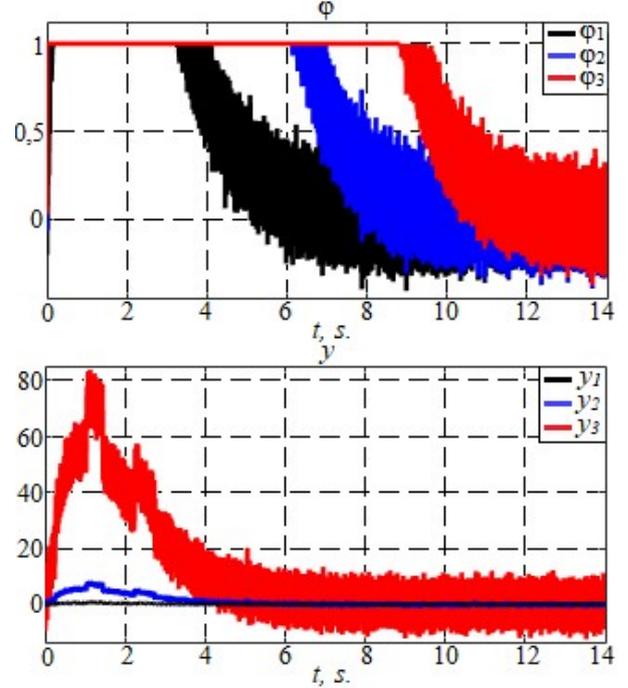

Fig. 6. Transient curves of regressors $\varphi_j$ and functions $y_j$.

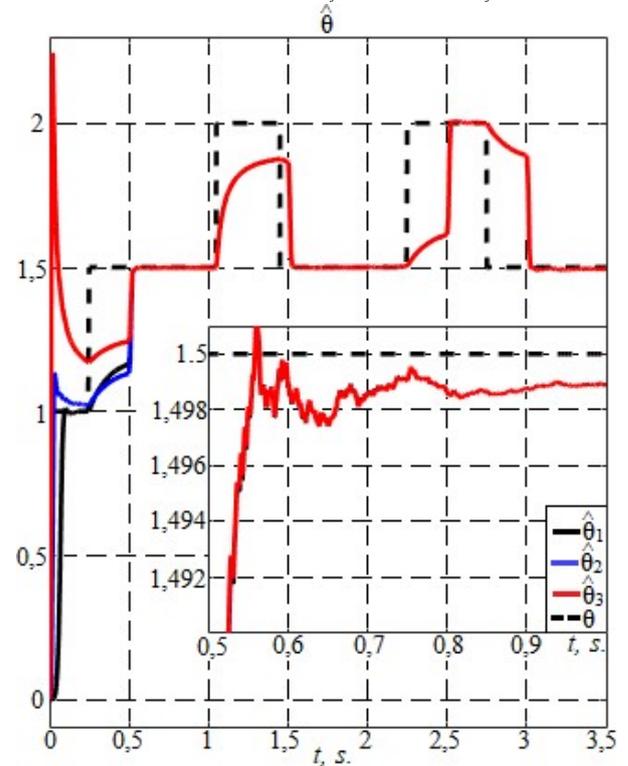

Fig. 7. Transient curves of estimates $\hat{\theta}_j$ under condition of disturbances.

As follows from Fig. 7, the parameter error $\tilde{\Theta} \in IB$ or $\tilde{\Theta} \in EUB$ over the whole experiment.

Thus, the conducted numerical simulation of the developed estimation law with normalization of the regressor excitation and interval-based integral filtering allowed us to validate the results of the theoretical part of the research.

## VII. CONCLUSION

In this research, the new method, which is based on the normalization procedure of the scalar regressor excitation [19] and integral filtering with exponential forgetting and resetting [15], has been proposed to estimate the piecewise-constant LRE parameters.

The developed approach does not require to meet the condition of the regressor persistent excitation and, when the regressor is finitely exciting, allows one to provide the exponentially bounded error of estimation of piecewise-constant LRE parameters. In addition, the presented estimation law provides the same parameter error convergence rate for regressors with various values of the amplitude (degree of excitation), which is an important result from the practical point of view.

In further research, it is planned to use the developed estimation law as a part of the CMRAC loop to adjust the controller parameters to improve the results obtained in [15].

## APPENDIX

### A. Proof of Theorem 1

To prove Theorem 1, $\Omega(t)$ and $\Upsilon(t)$ are rewritten over the interval $\left[t_k; t_{k+1}\right)$ as:

$$\Omega(t) = \underbrace{\int_{t_k}^{t_i} \exp\left(-\int_0^t \sigma d\tau_1\right) \varphi^2(\tau) d\tau}_{\Omega_1} + \underbrace{\int_{t_i}^{t_{k+1}} \exp\left(-\int_0^t \sigma d\tau_1\right) \varphi^2(\tau) d\tau}_{\Omega_2};$$

$$\Upsilon(t) = \underbrace{\int_{t_k}^{t_i} \exp\left(-\int_0^t \sigma d\tau_1\right) Y(\tau) \varphi(\tau) d\tau}_{\Upsilon_1} + \underbrace{\int_{t_i}^{t_{k+1}} \exp\left(-\int_0^t \sigma d\tau_1\right) Y(\tau) \varphi(\tau) d\tau}_{\Upsilon_2};$$

(A1)

With regard to (4), The unknown parameters are defined as follows over the interval $\left[t_k; t_{k+1}\right)$:

$$\Theta = \begin{cases} \Theta_0, \ \forall t \in [t_k; t_i) \\ \Theta_0 + \sum_j \theta_j h(t - t_j), \ \forall t \in [t_i; t_{k+1}) \end{cases};$$

(A2)

Combining (A1) and (A2), $\forall t \in [t_i; t_{k+1})$ the following regression is obtained:

$$\Upsilon = (\Omega_1 + \Omega_2)\Theta_0 + \Omega_2\Theta_1 = \Omega\Theta_0 + \Omega_2\Theta_1 = \Omega\left(-\tilde{\Theta} - \Theta_1 + \hat{\Theta}\right) + \Omega_2\Theta_1,$$

(A3)

where $\Omega_2$ is the unmeasurable regressor, $\Theta_1$ is the part of the vector of the unknown parameters.

Analyzing the stability of the law (28), the Lyapunov function is chosen as:

$$V = \frac{1}{2}\tilde{\Theta}^T\tilde{\Theta} = \frac{1}{2}\left\|\tilde{\Theta}\right\|^2$$

(A4)

The derivative of the quadratic form (A4), considering the equations (28) and (A3), is obtained as:

$$\dot{V} = \tilde{\Theta}^T\dot{\tilde{\Theta}} = \tilde{\Theta}^T\left[-\gamma\Omega\left(\Omega\hat{\Theta} - \Omega\left(-\tilde{\Theta} - \Theta_1 + \hat{\Theta}\right) - \Omega_2\Theta_1\right)\right] = \tilde{\Theta}^T\left[-\gamma\Omega\left(\Omega\tilde{\Theta} + \Theta_1\right) - \Omega_2\Theta_1\right)\right] = \tilde{\Theta}^T\left[-\gamma\Omega\left(\Omega\tilde{\Theta} + \Omega_1\Theta_1\right)\right]$$

(A5)

The upper bound of the regressor $\Omega_1$ is defined as its value at the time point $t_i$ for further analysis:

$$\Omega_1 = \int_{t_k}^{t_i} \exp\left(-\int_0^t \sigma d\tau_1\right) \varphi^2(\tau) d\tau \leq \Omega_1(t_i) = \Omega_{1UB}$$

(A6)

Considering (A6) and Proposition 6, the upper bound of the derivative (A5) $\forall t \in \left[\overline{T}_{0k}; t_{k+1}\right)$ is written as:

$$\dot{V} \leq -\gamma\Omega^2\left(\overline{T}_{0k}\right)\left\|\tilde{\Theta}\right\|^2 + \gamma T\Omega_{1UB}\|\Theta_1\|\left\|\tilde{\Theta}\right\|$$

(A7)

Let an axillary equation be introduced for some $a>0$ and $b>0$:

$$-a^2 + ab = \frac{1}{2}\left[-a^2 - (a-b)^2 + b^2\right] \leq -\frac{1}{2}a^2 + \frac{1}{2}b^2$$

(A8)

Having applied (A8) to (A7), it is obtained:

$$\dot{V} \leq -\frac{\gamma\Omega^2\left(\overline{T}_{0k}\right)}{2}\left\|\tilde{\Theta}\right\|^2 + \frac{\gamma T^2\Omega_{1UB}^2}{2\Omega^2\left(\overline{T}_{0k}\right)}\|\Theta_1\|^2 =$$

$$= -\gamma\Omega^2\left(\overline{T}_{0k}\right)V + \frac{\gamma T^2\Omega_{1UB}^2}{2\Omega^2\left(\overline{T}_{0k}\right)}\|\Theta_1\|^2$$

(A9)

The differential equation (A9) is solved $\forall t \in \left[\overline{T}_{0k}; t_{k+1}\right)$:

$$V \leq e^{-\gamma\Omega^2\left(\overline{T}_{0k}\right)\left(t-\overline{T}_{0k}\right)}V\left(\overline{T}_{0k}\right) + \frac{T^2\Omega_{1UB}^2}{2\Omega^4\left(\overline{T}_{0k}\right)}\|\Theta_1\|^2$$

(A10)

With regard to the inequality $\sqrt{a+b} \leq \sqrt{a} + \sqrt{b}$ and the definition (A4), it is obtained from (A10):

$$\forall t \in \left[\overline{T}_{0k}; t_{k+1}\right): \ \left\|\tilde{\Theta}(t)\right\| \leq e^{-0.5\gamma\Omega^2\left(\overline{T}_{0k}\right)\left(t-\overline{T}_{0k}\right)}\left\|\tilde{\Theta}\left(\overline{T}_{0k}\right)\right\| + \frac{T\Omega_{1UB}}{\Omega^2\left(\overline{T}_{0k}\right)}\|\Theta_1\|$$

(A11)

Therefore, following Definition 3, it is concluded that $\tilde{\Theta}(t) \in IB \ \forall t \in \left[\overline{T}_{0k}; t_{k+1}\right)$, as was to be proved in the first part of the theorem.

According to the statement of the theorem, the unknown parameters are defined as $\Theta = \Theta_0 + \Theta_1 \ \forall t \in [t_{k+1}; t_{k+2})$. Then $\forall t \in [t_{k+1}; t_{k+2})$ the regression equation can be written as $\Upsilon = \Omega\Theta$. Considering (28) and Proposition 6, the upper bound of the derivative of the quadratic form (A4) is obtained $\forall t \in \left[\overline{T}_{0(k+1)}; t_{k+2}\right)$:

$$\dot{V} = \tilde{\Theta}^T\dot{\tilde{\Theta}} = -\gamma\Omega^2\tilde{\Theta}^T\tilde{\Theta} \leq -\gamma\Omega^2\left(\overline{T}_{0(k+1)}\right)\left\|\tilde{\Theta}\right\|^2 = -2\gamma\Omega^2\left(\overline{T}_{0(k+1)}\right)V$$

(A12)

Taking into consideration the definition of $V$, the equation (A12) is solved $\forall t \in \left[\overline{T}_{0(k+1)}; t_{k+2}\right)$:

$$\forall t \in \left[\overline{T}_{0(k+1)}; t_{k+2}\right): \ \left\|\tilde{\Theta}(t)\right\| \leq e^{-0.5\gamma\Omega^2\left(\overline{T}_{0(k+1)}\right)\left(t-\overline{T}_{0(k+1)}\right)}\left\|\tilde{\Theta}\left(\overline{T}_{0(k+1)}\right)\right\|$$

(A13)

Whence it follows that $\tilde{\Theta}(t) \in IEB \ \forall t \in \left[\overline{T}_{0(k+1)}; t_{k+2}\right)$, as was to be proved in the second part of the theorem.

### B. Proof of Theorem 2

It follows from the theorem requirements that the unknown parameters are defined as $\Theta = \Theta_0 \ \forall t \in [t_k; t_{k+1})$. So, the regression equation is written as $\Upsilon = \Omega\Theta$. Considering (28) and Proposition 6, the upper bound of the derivative of the quadratic form (A4) is obtained $\forall t \in \left[\overline{T}_{0k}; t_{k+1}\right)$:

$$\dot{V} = \tilde{\Theta}^T\dot{\tilde{\Theta}} = -\gamma\Omega^2\tilde{\Theta}^T\tilde{\Theta} \leq -\gamma\Omega^2\left(\overline{T}_{0k}\right)\left\|\tilde{\Theta}\right\|^2 = -2\gamma\Omega^2\left(\overline{T}_{0k}\right)V$$

(B1)

Considering the definition (A4) of $V$, the equation (B1) is solved $\forall t \in \left[\overline{T}_{0k}; t_{k+1}\right)$:

$$\forall t \in \left[\overline{T}_{0k}; t_{k+1}\right): \ \left\|\tilde{\Theta}(t)\right\| \leq e^{-\gamma\Omega^2\left(\overline{T}_{0k}\right)\left(t-\overline{T}_{0k}\right)}\left\|\tilde{\Theta}\left(\overline{T}_{0k}\right)\right\|$$

(B2)

Whence it follows that $\tilde{\Theta}(t) \in IEB \ \forall t \in \left[\overline{T}_{0k}; t_{k+1}\right)$, which proves the first part of the theorem. To prove the second part of the theorem, in a similar way to (A1)-(A3), $\Omega(t)$ and $\Upsilon(t)$ are written as:

$$\Omega(t) = \underbrace{\int_{t_k}^{t_i} \exp\left(-\int_0^t \sigma d\tau_1\right) \varphi^2(\tau) d\tau}_{\Omega_1} + \underbrace{\int_{t_i}^{t} \exp\left(-\int_0^t \sigma d\tau_1\right) \varphi^2(\tau) d\tau}_{\Omega_2};$$

$$\Upsilon(t) = \underbrace{\int_{t_k}^{t_i} \exp\left(-\int_0^t \sigma d\tau_1\right) Y(\tau) \varphi(\tau) d\tau}_{\Upsilon_1} + \underbrace{\int_{t_i}^{t} \exp\left(-\int_0^t \sigma d\tau_1\right) Y(\tau) \varphi(\tau) d\tau}_{\Upsilon_2};$$

(B3)

According to the theorem statement, the unknown parameters are defined as:

$$\Theta = \begin{cases} \Theta_0, \ \forall t \in \left[ T^+ ; \ t_i \right] \\ \Theta_0 + \underbrace{\sum_j \theta_j h(t - t_j)}_{\Theta_1}, \ \forall t \in \left[ t_i ; \ t \right] \end{cases} ; \tag{B4}$$

Combining equations (B3) and (B4), $\forall t \geq t_i$ the regression (A3) is obtained. The derivative of the quadratic form (A4) $\forall t \geq t_i$ is (A5). Similarly to (A6), the upper bound of the regressor $\Omega_1$ is defined as its value at the time point $t_i$ :

$$\Omega_1 = \int_{T^+}^{t_i} \exp\left( -\int_0^{\tau} \sigma d\tau_1 \right) \varphi^2(\tau) d\tau \leq \Omega_1(t_i) = \Omega_{1UB} \tag{B5}$$

With regard to Proposition 7 and the equation (B5), the upper bound of the derivative (A5) $\forall t \geq t_e$ is written as:

$$\dot{V} \leq -\gamma \overline{\Omega}_{LB}^2 \left\| \tilde{\Theta} \right\|^2 + \gamma \Omega_{UB} \Omega_{1UB} \left\| \Theta_1 \right\| \left\| \tilde{\Theta} \right\| \tag{B6}$$

Using (A8), it is obtained from (B6):

$$\dot{V} \leq -\frac{\gamma \overline{\Omega}_{LB}^2}{2} \left\| \tilde{\Theta} \right\|^2 + \frac{\gamma \Omega_{UB}^2 \Omega_{1UB}^2}{2\overline{\Omega}_{LB}^2} \left\| \Theta_1 \right\|^2 = -\gamma \overline{\Omega}_{LB}^2 V + \frac{\gamma \Omega_{UB}^2 \Omega_{1UB}^2}{2\overline{\Omega}_{LB}^2} \left\| \Theta_1 \right\|^2 \tag{B7}$$

The differential equation (B7) is solved $\forall t \geq t_e$ :

$$V \leq e^{-\gamma \overline{\Omega}_{LB}^2 (t - t_e)} V(t_e) + \frac{\Omega_{UB}^2 \Omega_{1UB}^2}{2\overline{\Omega}_{LB}^4} \left\| \Theta_1 \right\|^2 \tag{B8}$$

Taking into consideration $\sqrt{a + b} \leq \sqrt{a} + \sqrt{b}$ and the definition (A4) of $V$, it is obtained from (B8):

$$\forall t \geq t_e : \left\| \tilde{\Theta}(t) \right\| \leq e^{-0.5\gamma \overline{\Omega}_{LB}^2 (t - t_e)} \left\| \tilde{\Theta}(t_e) \right\| + \frac{\Omega_{UB} \Omega_{1UB}}{\overline{\Omega}_{LB}^2} \left\| \Theta_1 \right\| \tag{B9}$$

Whence it follows that $\tilde{\Theta}(t) \in$ EUB $\forall t \geq t_e$, which proves the second part of the theorem.

### C. Proof of Theorem 3

According to the theorem requirements, $\forall t \geq t_r^+ \ \Theta = \Theta_0$. Then the regression equation is written as $\Upsilon = \Omega \Theta$. In this case, the first part of Theorem 3 coincides with the first part of Theorem 2, which has already been proven in Appendix B.

To prove the second part of the theorem, taking into consideration (28) and $\Upsilon = \Omega \Theta$, the upper bound of the derivative of the quadratic form (A4) is written as follows $\forall t \geq t_e$ :

$$\dot{V} = \tilde{\Theta}^T \dot{\tilde{\Theta}} = -\gamma \Omega^2 \tilde{\Theta}^T \tilde{\Theta} \leq -\gamma \overline{\Omega}_{LB}^2 \left\| \tilde{\Theta} \right\|^2 = -2\gamma \overline{\Omega}_{LB}^2 V \tag{C1}$$

With regard to the definition of $V$, the equation (C1) is solved $\forall t \geq t_e$ :

$$\forall t \geq t_e : \left\| \tilde{\Theta}(t) \right\| \leq e^{-\gamma \overline{\Omega}_{LB}^2 (t - t_e)} \left\| \tilde{\Theta}(t_e) \right\| \tag{C2}$$

Whence it follows that $\tilde{\Theta}(t) \in$ GES $\forall t \geq t_e$, which was to be proved in the second part of the theorem.